\documentclass[
floatfix,
aps,
prb,
twocolumn,
superscriptaddress,
]{revtex4}

\usepackage{amssymb}
\usepackage{latexsym}
\usepackage[dvips]{graphicx}
\usepackage{amsmath}
\usepackage{graphicx}
\usepackage{dcolumn}
\usepackage{amsfonts}
\usepackage{bm}
\usepackage{epsfig}
\usepackage{times}

\newcommand{\be}{\begin{equation}}
\newcommand{\ee}{\end{equation}}
\newcommand{\bea}{\begin{eqnarray}}
\newcommand{\eea}{\end{eqnarray}}

\newcommand{\F}{\mathcal F}
\newcommand{\pc}{{\mathcal P}}

\def\pst{{\mathcal P}^\star}

\def\Eac{\mathcal{E}_{\mathrm{ac}}}

\def\eac{\epsilon}

\def\pac{\delta}

\def\esm{\eta_{\it sm}}
\def\esh{\eta_{\it sh}}
\def\ein{\eta_{\it in}}
\def\pstin{\pst_{\it in}}
\def\pstsm{\pst_{\it sm}}
\def\pstsh{\pst_{\it sh}}

\def\epseff{\epsilon_{\mathrm{eff}}}

\def\oc{\omega_{\mbox{\scriptsize {c}}}}

\def\tq{\tau_0}
\def\ttr{\tau}
\def\tem{\tau_{\it em}}
\def\tsh{\tau_{\it sh}}
\def\tsm{\tau_{\it sm}}

\def\tin{\tau_{\it in}}

\def\bo{\beta_{\omega}}

\newcommand{\req}[1]{Eq.\,(\ref{#1})}
\newcommand{\rEq}[1]{Equation\,(\ref{#1})}

\newcommand{\rfig}[1]{Fig.\,\ref{#1}}

\newcommand{\rref}[1]{Ref.\,\onlinecite{#1}}

\newcommand{\rrrefs}[3]{Refs.\,\onlinecite{#1},\,\onlinecite{#2},\,\onlinecite{#3}}

\begin{document}
\title{
Multiphoton microwave photoresistance in a high-mobility two-dimensional electron gas
}

\author{A.\,T. Hatke}
\affiliation{School of Physics and Astronomy, University of Minnesota, Minneapolis, Minnesota 55455, USA}

\author{M.\,Khodas}
\affiliation{Department of Physics and Astronomy, University of Iowa, Iowa City, Iowa 52242, USA}

\author{M.\,A. Zudov}
\affiliation{School of Physics and Astronomy, University of Minnesota, Minneapolis, Minnesota 55455, USA}

\author{L.\,N. Pfeiffer}
\affiliation{Princeton University, Department of Electrical Engineering, Princeton, New Jersey 08544, USA}

\author{K.\,W. West}
\affiliation{Princeton University, Department of Electrical Engineering, Princeton, New Jersey 08544, USA}

%\received{10 October 2011}

\begin{abstract}
We report on experimental and theoretical studies of microwave-induced resistance oscillations in a two-dimensional electron gas over a wide range of microwave intensities.
We observe a distinct crossover from linear to sublinear power dependence of the oscillation amplitude and a concomitant narrowing of the oscillation extrema.
To explain our observations we propose a theory based on the quantum kinetic equation at arbitrary microwave power. 
Taken together, these findings demonstrate a crucial role of multiphoton processes at elevated microwave intensities.
\end{abstract}
%\pacs{73.43.Qt, 73.63.Hs, 73.21.-b, 73.40.-c}
\maketitle

Microwave-induced resistance oscillations (MIRO) appearing in very high Landau levels of high-mobility two-dimensional electron gas (2DEG) have been discovered more than a decade ago.\cite{zudov:2001a,ye:2001}
Since then, a variety of other remarkable low-field transport phenomena have been identified and actively studied.
Among these are phonon-induced\cite{zudov:2001b,raichev:2009,hatke:2009b,hatke:2011d} and Hall field-induced\cite{yang:2002,zhang:2007a,vavilov:2007,hatke:2009c,hatke:2011a}
resistance oscillations, several classes of ``combined'' oscillations,\cite{zhang:2007c,zhang:2008,hatke:2008a,hatke:2008b,khodas:2008,khodas:2010,dmitriev:2010b} zero-resistance states,\cite{mani:2002,zudov:2003,willett:2004,yang:2003,andreev:2003,smet:2005,auerbach:2005,zudov:2006a,zudov:2006b,wiedmann:2010b,dorozhkin:2011} several distinct zero-differential resistance states,\cite{zhang:2007c,zhang:2008,hatke:2010a,bykov:2007,wiedmann:2011a} and, most recently, a very sharp and strong photoresistivity peak near the second harmonic of the cyclotron resonance. \cite{dai:2010,hatke:2011b,hatke:2011c}
Microwave-induced conductance oscillations and zero-conductance states have also been realized for electrons on liquid helium.\cite{konstantinov:2009b,konstantinov:2010}

Despite a huge body of experimental\cite{dorozhkin:2003,zudov:2004,dorozhkin:2005,studenikin:2005,studenikin:2007,mani:2004e,mani:2005,yang:2006,bykov:2006,zhang:2008b,zudov:2009,fedorych:2010,tung:2009,hatke:2009a,andreev:2011} and theoretical\cite{ryzhii:1970,ryzhii:1986,ryzhii:2003a,ryzhii:2003d,ryzhii:2004,durst:2003,lei:2003,shi:2003,dmitriev:2003,vavilov:2004,dmitriev:2005,auerbach:2007,dmitriev:2007b,dmitriev:2007,dmitriev:2009b} work on microwave photoresistance, there are still many unsolved puzzles, such as activated temperature dependence at the oscillation minima,\cite{mani:2002,zudov:2003,willett:2004} the role of microwave polarization,\cite{smet:2005} and the effect of an in-plane magnetic field.\cite{mani:2005,yang:2006}
Another outstanding issue, which is the main subject of this Rapid Communication, is the dependence on microwave intensity which to date was not investigated in detail.

MIRO can originate from either the {\em displacement} mechanism,\citep{ryzhii:1970,ryzhii:1986,durst:2003,lei:2003,vavilov:2004} stepping from the modification of impurity scattering by microwaves, or the {\em inelastic} mechanism,\citep{dorozhkin:2003,dmitriev:2003,dmitriev:2005,dmitriev:2007b,dmitriev:2007,dmitriev:2009b} owing to the microwave-induced non-equilibrium distribution of electrons.
In the most widely-used approximation of {\em weak} microwave intensities, {\em both} mechanisms lead to a simple expression for the photoresistivity,
\be
\frac {\delta \rho_\omega}{\rho_D} \simeq - 4\lambda^2 \pi \eac A(\pc)  \sin 2\pi\eac,
\label{eq.miro}
\ee
where $\rho_D$ is the Drude resistivity, $\eac = \omega/\oc$, $\omega=2\pi f$ is the microwave frequency, $\oc=e B /m^*$ is the cyclotron frequency of an electron with an effective mass $m^*$, $\lambda = \exp(-\pi/\oc\tq)$ is the Dingle factor, $\tq$ is the quantum lifetime, $A(\pc)=\eta \pc$ is the reduced amplitude,  $\eta$ is mechanism-dependent scattering parameter, and $\pc$ is the dimensionless microwave power which, for active circular polarization, is given by\cite{khodas:2010}
\be
\pc(\eac)= 
\frac{\pc^{0}}{(1 - \eac^{-1})^2+\bo^2},~\pc^{0}=\frac{e^2\Eac^2v_F^2}{\epseff \hbar^2 \omega^4}\,.
\label{pc2}
\ee
Here, $\bo\equiv(\omega\tem)^{-1}$, $\tem^{-1}=n_e e^2/2\sqrt{\epseff}\epsilon_0m^*c$, $2\sqrt{\epseff}=\sqrt{\varepsilon}+1$, $\varepsilon=12.8$ is the dielectric constant of GaAs, $v_F$ is the Fermi velocity, and $\Eac$ is the external microwave field. 
\rEq{eq.miro} thus predicts linear power dependence of the oscillation amplitude and a power-independent {\em phase}, $\delta\equiv n-\eac^+=1/4$, where $n$ is an integer and $\eac^+=\eac$ at the closest maximum.

Experimentally, however, all studies,\cite{ye:2001,studenikin:2004,willett:2004,mani:2004a,mani:2010,inarrea:2010} except \rref{zudov:2003}, found strongly sublinear power dependence.\cite{note:1}
These reports indicate that \req{eq.miro}, while used widely to analyze and interpret the experimental data, is {\em not} valid under typical experimental conditions. 
At the same time, theoretical studies of the high intensity regime have only considered the limit of smooth disorder.\cite{vavilov:2004,dmitriev:2005,dmitriev:2007,note:2}
However, it is well established both theoretically and experimentally that in a realistic high-mobility 2DEG sharp disorder plays a crucial role in many of the above mentioned non-equilibrium phenomena,\cite{yang:2002,zhang:2007a,zhang:2007c,zhang:2008,hatke:2008a,vavilov:2007,hatke:2008b,hatke:2009c,hatke:2011a,khodas:2008,khodas:2010} including MIRO.\cite{hatke:2009a,dmitriev:2009b} 

In this Rapid Communication we report on experimental and theoretical studies of microwave photoresistance in a very high-mobility 2DEG performed over a three orders of magnitude range of microwave intensities.
We observe a distinct crossover from linear to sublinear power dependence of the oscillation amplitude.
This crossover is accompanied by narrowing of the oscillation extrema, which is reflected as a decreasing phase.
To explain our observations we propose a theory based on the quantum kinetic equation. 
Applicable at {\em arbitrary} microwave power, our model takes into account both the displacement and the inelastic contributions within a model of {\em mixed} disorder.
Taken together, the findings demonstrate an important role of {\em multiphoton processes} at elevated microwave power and should be useful in interpreting prior experiments.\cite{ye:2001,studenikin:2004,willett:2004,mani:2004a,mani:2010}

Our Hall bar sample was fabricated from a symmetrically doped GaAs/AlGaAs quantum well with density $n_e =2.9 \times 10^{11}$ cm$^{-2}$ and mobility $\mu = 2.4 \times 10^7$ cm$^2$/Vs.
The microwave radiation of frequency $f=33$ GHz was delivered to the sample using a WR-28 waveguide.
Resistance was measured at $T\simeq 1.5$ K using a standard low frequency lock-in technique.
Microwave intensity was varied by a calibrated attenuator.
Maximum power at the end of the waveguide is estimated as $P_0 \approx 1$ mW.

%%%%%%%%%%%%%%%%%%%%%%%%%%%%%%%%%%%%%%%%%%%%%%%%%
%fig 1
\begin{figure}[t]
\includegraphics{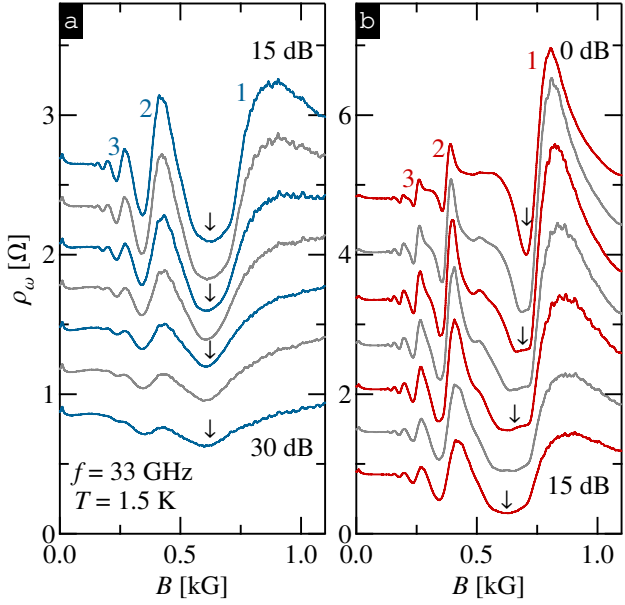}
\vspace{-0.1 in}
\caption{(Color online) Normalized microwave magnetoresistivity $\rho_\omega(B)$ measured at attenuations (a) from 30 dB to 15 dB  and (b) from 15 to 0 dB, in a step of 2.5 dB.
Integers mark MIRO order.
}
\vspace{-0.15 in}
\label{fig1}
\end{figure}
%%%%%%%%%%%%%%%%%%%%%%%%%%%%%%%%%%%%%%%%%%%%%%%%%
In \rfig{fig1} we present microwave magnetoresistivity $\rho_\omega(B)$ measured at attenuations from 30 dB to 15 dB [Fig.\,\ref{fig1}(a)] and from 15 to 0 dB [Fig.\,\ref{fig1}(b)], in a step of 2.5 dB.
At low intensities [Fig.\,\ref{fig1}(a)] the MIRO amplitude grows roughly linearly (5 dB is equivalent to a factor of about 3) and the positions of the maxima/minima remain constant (cf.\,$\downarrow$), in agreement with \req{eq.miro}.
At higher intensities [Fig.\,\ref{fig1}(b)] the amplitude grows considerably slower and eventually starts to decrease, likely as a result of heating.\cite{note:3}
At the same time, the extrema become sharper while moving towards the closest cyclotron resonance harmonic (cf.\,$\downarrow$) indicating considerable reduction of the phase.
These observations confirm that at high intensities microwave photoresistance can no longer be described by \req{eq.miro}.

%%%%%%%%%%%%%%%%%%%%%%%%%%%%%%%%%%%%%%%%%%%%%%%%%
%fig 2
\begin{figure}[t]
\includegraphics{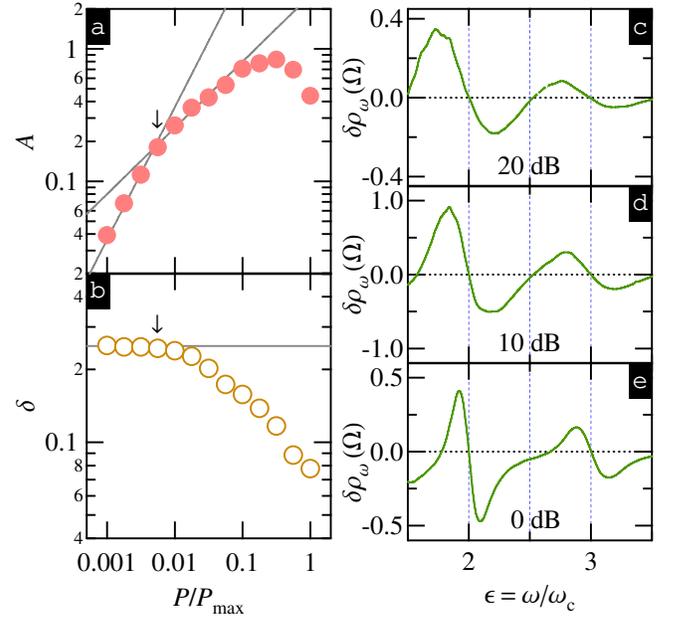}
\vspace{-0.1 in}
\caption{(Color online)
(a) Amplitude $A$ and (b) phase $\pac$ measured at the second MIRO maximum versus normalized microwave intensity $P/P_{\max}$.
Solid lines in (a) represent linear and square-root dependences (see text). 
Photoresistivity $\delta\rho_\omega (\eac)$ at attenuation of (c) 20 dB, (d) 10 dB, and (e) 0 dB.
}
\vspace{-0.15 in}
\label{fig2}
\end{figure}
%%%%%%%%%%%%%%%%%%%%%%%%%%%%%%%%%%%%%%%%%%%%%%%%%

Using the data in \rfig{fig1} we extract the amplitude $A$ and the phase $\pac$ for $n=2$ and present the results in \rfig{fig2}(a) and \rfig{fig2}(b), respectively, as a function of microwave power normalized to its maximum value, $P/P_{\max}$.
Plotted in such a way, the data clearly demonstrate the existence of two distinct regimes.
At low intensities, the amplitude grows roughly linearly with power and the phase is power independent.
Above a certain intensity, marked by $\downarrow$, the amplitude becomes strongly sublinear and, simultaneously, the phase starts to decrease.

The crossover between the regimes of low and high intensities can also be established by examining the evolution of the oscillation waveform.
Indeed, as illustrated in \rfig{fig2}(c)-2(e) showing photoresistivity $\delta\rho_\omega (\eac)$ at 20 dB, 10 dB, and 0 dB, respectively, the waveform undergoes dramatic changes.
At low intensity [Fig.\,\ref{fig2}(c)] the oscillations are well described by a damped sinusoidal, in agreement with \req{eq.miro}.
However, already at intermediate power [Fig.\,\ref{fig2}(d)], one clearly sees that the extrema are pushed towards integer $\eac$. 
At maximum power [\rfig{fig2}(e)], the extrema become very sharp reflecting dramatically reduced phase. 

To explain the observed crossover we construct a theory based on quantum kinetics approach \cite{vavilov:2004,dmitriev:2005,khodas:2008} for arbitrary $\pc$.
Our main result is given by
\begin{align} \label{c5a}
\frac{ \delta \rho_\omega }{  \rho_{D} } = 2 \lambda^{2} \left [\F(\eac) -1 \right],
\end{align}
where, for circular polarization,
\begin{align} \label{f}
\frac{\F(\eac) }{ \ttr } = 
( \eac \bar{\gamma}(\xi))' + \frac{  2 \eac \bar{\gamma}(\xi) \gamma(\xi)' }{ \tin^{-1} + \tau_{0}^{-1}  - \gamma(\xi)},
\end{align}
$\xi = 2 \sqrt \pc \sin \pi \eac $, $\ttr^{-1}$  ($\tin^{-1}$) is the transport (inelastic) scattering rate, and the prime denotes the derivative with respect to $\eac$.

For the model of mixed disorder\cite{vavilov:2007} with sharp (smooth) disorder scattering rate $\tsh^{-1}$ ($\tsm^{-1}$) and a typical scattering angle off smooth disorder $\chi^{1/2}\ll 1$, we find
\begin{subequations}\label{gammainmodel=}
\bea
\label{DisModel17}
\gamma(\xi) &=&  \frac{ J_0^2(\xi) }\tsh+ \frac 1 \tsm \frac 1 { (1 + \chi \xi^2)^{1/2}} \,,\\
\label{DisModel20} 
\bar{\gamma}(\xi) &=& \frac{ 1 }{ \tsh }
\left[ J_0^2(\xi) - J_1^2(\xi) \right]
 +
\frac{ \chi }{\tsm } \frac{ 1 - \chi \xi^2 / 2}{(1 + \chi
\xi^2)^{5/2}}.
\eea
\end{subequations}
In the limit of smooth disorder, $\tsh^{-1}=0$, the first term in \req{f} reproduces the displacement contribution obtained in \rref{vavilov:2004} while the second term agrees with the inelastic contribution obtained in \rref{dmitriev:2005} assuming $\pc \ll \ttr/\tq$.
However, the real situation in the smooth disorder limit is believed to be more complicated.
Already in the regime linear in microwave power, contributions from additional quadrupole and photovoltaic mechanisms, due to the excitation of second angular and first temporal harmonics of the distribution functions, respectively, become relevant,\cite{dmitriev:2007,dmitriev:2009b} and at higher power the number of important angular and temporal harmonics increases.\cite{dmitriev:2007}

In this work we consider an experimentally relevant situation when the disorder contains both smooth and sharp components. 
The latter is assumed strong enough so that its contribution to the transport relaxation rate is comparable to that of the smooth disorder, $\tsh^{-1} \gtrsim \chi\tsm^{-1}$.
In this case the quadrupole and photovoltaic mechanisms become negligible\cite{note:5} while the displacement and the inelastic contributions can be separated [see Eq.~\eqref{f}].
%%%%%%%%%%%%%%%%%%%%%%%%%%%%%%%%%%%%%%%%%%%%%%%%%
%fig 3
\begin{figure}[t]
\includegraphics{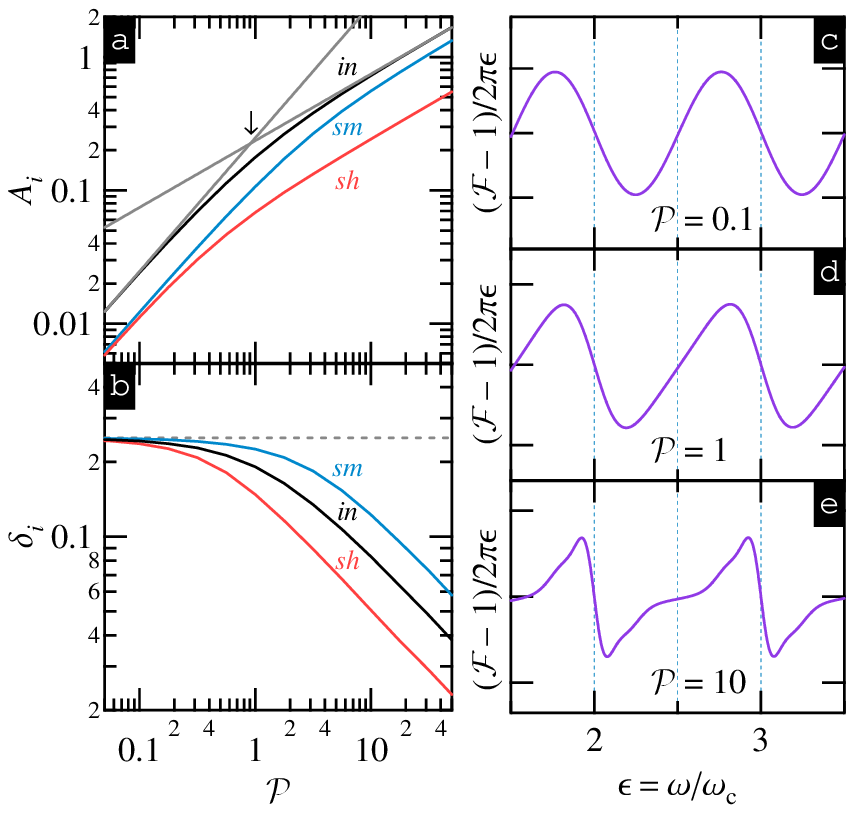}
\caption{(Color online)
(a) Amplitudes $A_i(\pc)$ and (b) phases $\pac_i(\pc)$ calculated using \req{c5a} with $\chi=0.022$, $\tq=20$ ps, $\ttr = 0.86$ ns, and  $\tin = 0.10$ ns.
Asymptotes (lines) cross at $\pstin$.
$[\F(\eac)-1]/2\pi\eac$ vs $\eac$ calculated for (c) $\pc = 0.1$, (d) $\pc=1$, and (e) $\pc=10$.
}
\vspace{-0.15 in}
\label{fig3}
\end{figure}
%%%%%%%%%%%%%%%%%%%%%%%%%%%%%%%%%%%%%%%%%%%%%%%%%
We analyze our results in terms of the reduced amplitude,
\be
A(\pc) = \frac {\F(\eac^+)-1}{2\pi\eac^+}\,,
\label{ampl}
\ee
and introduce indicies $i={\it sm,sh,in}$ to describe smooth displacement, sharp displacement, and inelastic contributions, respectively.
At low $\pc$, we find, in agreement with \req{eq.miro} and 
\rrrefs{vavilov:2004}{dmitriev:2005}{dmitriev:2009b},
\be
A_i(\pc)\simeq \eta_i \pc,~~\pac_i(\pc) \simeq 1/4,
\ee
where $\esm=6\chi^2 \ttr/\tsm$, $\esh=3\ttr/2 \tsh$, and $\ein=2\tin/\ttr$.
In the limit of high $\pc$, {\em all} contributions show a square-root (inverse square-root) dependence of the amplitude (phase),
\be
A_i(\pc) \simeq \eta_i (\pst_i\pc)^{1/2},~~\pac_i(\pc) \simeq (\pst_i/\pc)^{1/2}\,.
\ee
Here, $\pstsm \simeq 0.05/\chi$, $\pstsh \simeq 0.4$, $\pstin \simeq \min \left \{1,\,\ttr/2\tin \right \}$.\cite{note:4} 

Calculated dependences of the amplitudes, $A_i$, and phases, $\delta_i$, on microwave intensity $\pc$ are shown in \rfig{fig3}(a) and 3(b), respectively.
Distinct crossovers in both $A_i$ and $\delta_i$ dependences are evident for all three contributions.
In \rfig{fig3}(c)-3(e) we present $[\F(\eac)-1]/2\pi\eac$, containing {\em all} contributions, as a function of $\eac$ for (c) $\pc = 0.1$, (d) $\pc=1$, and (e) $\pc=10$ demonstrating considerable reduction of the phase at higher $\pc$, much the same as in the experiment, \rfig{fig2}(c)-2(e).
While our data is in a good qualitative agreement with the theory, the observed crossover is noticeably sharper than theoretically predicted [cf.\,\rfig{fig3}(a) and \rfig{fig2}(a)].
Possible reasons for this discrepancy are heating, uncertain polarization, and the inhomogeneity of microwave field, which are not included in our theory.
 
We now illustrate the appearance of the crossover for the sharp displacement contribution given by the first terms in $\F$, \req{f} and in $\bar\gamma(\xi)$, \req{DisModel20}.
Since $|\xi(\eac)| \le 2 \sqrt{\pc}$, at low intensity we can use the Taylor expansion of $\bar{\gamma}(\xi)$ leading to Eq.~\eqref{eq.miro}. 
At high intensity, the variable $\xi(\eac)$ spans an interval wide enough for the amplitude to be determined by the first maximum of the derivative $d\bar\gamma(\xi)/d\xi$. 
As the first term in Eq.~\eqref{DisModel20} involves Bessel functions, its variation scale is of the order of unity.
For that reason the maximum occurs at $\xi^+\equiv\xi(\eac^+) \simeq 1$ yielding $\eac^+ \simeq n - 1/2\pi\sqrt\pc$.
The amplitude then scales as $d\xi^+/d\eac^+ \approx 2 \pi \sqrt{\pc}$, entering the full derivative $d\bar{\gamma}(\xi^+)/d\epsilon^+$.
Explicit expression for the sharp disorder contribution is obtained substituting 
\eqref{DisModel20} with $\tsm^{-1} = 0$ to the first term of \eqref{f},
\begin{align}\label{disp:sh}
\F(\eac) =  & \frac{  \ttr }{ \tsh}
\Big\{ 
J_0^2(\xi) - J_1^2(\xi) 
\notag \\
& + 2 \pi \epsilon \sqrt{\pc} \cos(\pi \eac) J_1(\xi) \left[ J_2(\xi) - 3 J_0(\xi) \right]  
\Big\}.
\end{align}
Since the second line of Eq.~\eqref{disp:sh} contains a factor $2\pi\eac$, it dominates at $\eac \gtrsim 1$ and gives rise to the crossover as discussed above.
Similar considerations apply to smooth displacement contribution, given by the second term of \req{DisModel20}.

The inelastic contribution, given by the second term of \req{f}, represents the balance between the excitation (numerator) processes, driving the distribution function out of equilibrium, and its relaxation (denominator).
At $\tin \lesssim \ttr$, the crossover occurs at $\pst \simeq1$, very much like the sharp displacement contribution discussed above.
Indeed, the excitation rate reaches its maximum at $\xi=\xi^+ \simeq 1$ and, since $\gamma(\xi^+) \approx 1/ \tau_0$, the relaxation rate remains roughly constant, $\approx \tin^{-1}$.
In the opposite case of $\tin \gg \tau $, it is $\gamma(\xi)$ in the relaxation rate, varying in a narrow interval $|\xi(\eac)| \lesssim \sqrt{\ttr/\tin} \ll 1$, which controls the crossover.
As a result, $\xi^+ \ll 1$, and the theory of \rref{dmitriev:2005} is applicable with the result
\begin{align}\label{Finel}
\F(\eac)  = - \frac{4\pi \pc \eac \sin 2\pi\eac}{\ttr /\tin+2\pc \sin^2\pi \eac}\, .
\end{align}
The result \eqref{Finel} holds provided $\pc \sin^2(\pi\eac) \ll \ttr/\tq = \chi^{-1}$, $\chi \ll 1$ for the case of smooth disorder, $\tsh^{-1}=0$. 
However, when $\tsh^{-1} \gtrsim \chi\tsm^{-1}$ the applicability range of Eq.~\eqref{Finel} is reduced to $\pc \sin^2(\pi\eac) \ll 1$, as follows from Eqs.~\eqref{f} and \eqref{gammainmodel=}. 

Deviations from the linear power dependence of the numerator and the denominator are due to multiphoton processes and are captured by these equations.
The extremum of Eq.~\eqref{Finel} is reached at $\pc \sin^2(\pi\eac^{+}) = \ttr/2\tin$, and in the regime of slow inelastic relaxation $\tin \gg \ttr$ the one-photon approximation suffices to describe the maximum of MIRO adjacent to the cyclotron resonance harmonics.
Away from $\epsilon = n$ and at high intensities the inelastic contribution is strongly suppressed because each impurity scattering event contributes to the relaxation which saturates at $1/ \tq$, $\gamma(\xi) \approx 0$.
This occurs because of the multiphoton processes which we discuss below.

Observed crossovers reflect the transition from single- to multiphoton regime at elevated intensities. 
The rate of $N$-photon scattering process is $J_N^2 (\sqrt{ 2 \pc ( 1- \cos \theta ) }) / \tau_{\theta} $, where $J_N(x)$ is the Bessel function of the order $N$ and $\tau_{\theta}^{-1}$ is the scattering rate on angle $\theta$.\cite{khodas:2010}
Since $J_N(x)$ is strongly suppressed for $N \gtrsim x$, a typical number of participating photons is $N_i \approx \max \{1,\theta_i \sqrt{\pc }\}$, where $\theta_i$ is the contribution-dependent characteristic scattering angle. 
As a result, multiphoton processes become relevant at $\pc \gtrsim \pst_i \approx 1/ \theta_i^2$.
Since the scattering angles off smooth and sharp disorders are parametrically different, so are the above crossovers.

The square-root scaling of the amplitude, $A_i\propto N_i \propto \sqrt{\pc}$, at $\pc \gtrsim \pst_i$, is naturally explained by noting that the fraction of electrons which can absorb or emit $N_i$-photons is given by $N_i \hbar\omega/\varepsilon_F$, where $\varepsilon_F$ is the Fermi energy.
The inverse square-root dependence of the phase $\pac_i = (4N_i)^{-1} \propto \pc^{-1/2}$ follows directly from $N_i \eac^+ = n - 1/4$.

In summary, our experimental and theoretical studies of microwave photoresistance of a very high-mobility 2DEG over a wide range of microwave intensities revealed two distinct regimes. 
In one, low intensity regime, the oscillation amplitude grows linearly with power while the phase remains power independent. 
In the opposite, high intensity regime, the amplitude exhibits square-root dependence of the amplitude and the inverse square-root dependence of the phase.
The details of the crossover between these two regimes depends on the ratio $\tin/\ttr$.
For slow inelastic relaxation, $\tin/\ttr \gtrsim 1$ the crossover is described by Eq.~\eqref{Finel}.\cite{note:6}
For $\tin/\ttr \lesssim 1$ the crossover  occurs when a typical number of emitted (absorbed) photons becomes large, and is governed by multiphoton processes.
Since the corresponding crossover scale depends on the type of disorder and the transport mechanism it can be potentially useful in separating different contributions.
Taken together, our findings demonstrate an important role of multiphoton processes at elevated microwave power helping to understand prior experiments \cite{ye:2001,studenikin:2004,willett:2004,mani:2004a,mani:2010} reporting sublinear power dependence of the oscillation amplitude.

We thank I. Dmitriev and M. Dyakonov for discussions.
The work at Minnesota was supported by NSF Grant No. DMR-0548014.
The work at Princeton was partially funded by the Gordon and Betty Moore Foundation as well as the NSF MRSEC Program through the Princeton Center for Complex Materials (DMR-0819860).
M.K. acknowledges support from University of Iowa.


\begin{thebibliography}{77}
\bibitem[{not({\natexlab{c}})}]{note:3}
\bibinfo{note}{The growth of the zero-field resistivity indicates an increase
  of electron temperature by about 1 K from 5 dB to 0 dB. Due to such heating
  (and relatively low microwave frequency) zero-resistance states are not
  observed.}
\bibitem[{not({\natexlab{e}})}]{note:4}
\bibinfo{note}{Using $\chi=0.022$, $\tq=20$ ps, $\ttr = 0.86$ ns, and $\tin =
  0.10$ ns, we estimate $\pstin \simeq 1$, $\pstsm \simeq 2.4$.}
\bibitem[{not({\natexlab{f}})}]{note:6}
\bibinfo{note}{In this case the temperature dependence of MIRO amplitude is
  expected to cross over from quadratic, $\propto T^2$ to linear, $\propto T$.}
\end{thebibliography}
\end{document}